# Near total magnetic moment compensation without reduction in $T_C$ of $Mn_2V_{0.5}Co_{0.5}Z$ (Z=Ga,Al) Heusler compounds

**Midhunlal P V[1], Arout Chelvane J[2], Arjun Krishnan U M[3] and Harish Kumar N[1]**

1. Department of Physics, Indian Institute of Technology-Madras, Chennai-600036, India.
2. Defense Metallurgical Research Laboratory, Kanchanbagh (PO), Hyderabad, India.
3. School of Physical Sciences, National Institute of Science Education and Research (NISER), HBNI, Bhubaneswar-752050, Odisha, India

Corresponding Author's E-mail id: nhk@iitm.ac.in

**Abstract :** $Mn_2V_{1-x}Co_xZ$ (Z=Ga,Al and x=0, 0.25, 0.5, 0.75, 1) Heusler compounds have been synthesized to investigate the effect of Co substitution at the V site on the magnetic moment and Curie temperatures of half-metallic ferrimagnets $Mn_2VGa$ and $Mn_2VAl$. The Co substituted compounds show a non linear decrease in lattice parameter without altering the crystal structure of the parent compounds. The end members $Mn_2VGa$ and $Mn_2CoGa$ have the saturation magnetization of 1.80 $\mu_B$/f.u and 2.05 $\mu_B$/f.u respectively whereas for the $Mn_2V_{0.5}Co_{0.5}Ga$ compound, a near total magnetic moment compensation (0.10 $\mu_B$/f.u ) was observed due to the ferrimagnetic coupling of Mn with parallelly aligned V and Co. The Co substituted $Mn_2VAl$ has also shown the similar trend with compensated magnetic moment value of 0.06 $\mu_B$/f.u for x=0.5. The Curie temperatures of the compounds including the x=0.5 composition are well above the room temperature (more than 700 K) which is in sharp contrast to the earlier reported values of 171 K for the (MnCo)VGa and 105 K for the (MnCo)VAl compounds (substitution at the Mn site). The magnetic moment compensation without significant reduction in $T_C$ indicates that the V site substitution of Co does not weaken the magnetic interaction in $Mn_2VZ$ (Z=Ga, Al) compounds which is in contrary to the earlier experimental reports on Mn site substitution.

## 1. Introduction

Half-metals are considered to be ideal spin injection electrode materials in magnetic multilayer devices such as spin valves and MRAMs as they have 100% spin polarized charge carriers at the Fermi level. Among the various half-metallic materials such as transition metal oxides (eg: $CrO_2$, $Fe_3O_4$), dilute magnetic semiconductors (eg: Mn doped GaAs), Perovskite manganites (eg: $La_{0.7}Sr_{0.3}MnO_3$) and metal chalcogenides (eg: CrSe), Heusler compounds have got much attention due to their high Curie temperature and matching lattice parameter with the commonly used semiconducting substrates which are desirable for spintronic applications [1-3]. But the stray magnetic field produced by these magnetic materials may affect the performance of the components in a spintronic circuit. Hence it is desirable to have a half-metal with zero magnetic moment as the ideal spin injection electrode material. Usually antiferromagnets have zero magnetic moment as they have equal number of spin up and spin down electrons but are not half-metallic. Half-metallic Heusler compounds obeys Slater-Pauling relation which relates the total spin magnetic moment per unit cell ($M_t$ in $\mu_B$) with the total number of valance electrons ($Z_t$). For full-Heusler compounds, it can be expressed as $M_t=Z_t-24$ and for the half-Heusler compounds it turns out to be $M_t=Z_t-18$ [4]. This implies that the half-metallic full-Heusler compounds with 24 valance electrons and half-Heusler compounds with 18 valance electrons should exhibit zero magnetic moment. These materials with 100 % spin polarization at the Fermi level and zero magnetic moment is referred to as half-metallic antiferromagnets or more accurately half-metallic fully compensated ferrimagnets (HMFCFs). Hence HMFCFs can be utilized as spin injection electrode and as a material for the magnetic tip in spin polarized scanning tunneling microscope (SPSTM) (since it does not alter the magnetic behavior of the sample) [5]. Here it is to be mentioned that in a normal antiferromagnet, compensation of magnetic moments happens due to the same kind of atoms sitting at different atomic sites (like Mn in MnO). But in the case of $C1_b$ type (half-Heusler compounds, XYZ) HMFCFs, compensation happens between the two different atoms with the same site symmetry as in the case of MnCrSb whereas for the $D0_3$ type compounds, (full-Heusler compounds with X=Y) compensation happens between the same kind of atoms with different site symmetry as in the case of $Mn_3Ga$ [6]. The first predicted HMFCF was MnCrSb by de Groot in 1991 [5] and thereafter several materials such as $V_7MnFe_8Sb_7In$ [7], $Cr_2CoGa$ [8], $Cr_2MnZ$ (Z=P, As, Sb, Bi) [9], CrVXAl (X= Ti, Zr, Hf) [10], $Mn_2Z$ (Z=Si, Ge) [11], $Mn_3Ga$ [6] and $Mn_{2.41}Pt_{0.59}Ga$ [12] are predicted to have HMFCF property on the basis of theoretical calculations. But none of these compounds were successfully synthesized with the desired crystal structure and zero moment half-metallic property. But recent report on $D0_3$ $Mn_3Al$ films shows almost compensated ferrimagnetism with saturation magnetization ($M_S$) of 0.11±0.04 $\mu_B$ and $T_C$=605 K [13]. The tetragonal $D0_{22}$ $Mn_3Ga$ bulk compound exhibits a low $M_S$ of 0.25 $\mu_B$ and a high $T_C$ of 730 K [14]. Another interesting material in the HMFCF family is Rh added cubic $Mn_2Ga$ film with $C1_b$ structure. $Mn_2Ru_{0.5}Ga$ compound with 21 valance electrons was identified as zero-moment ferrimagnet [15]. I. Galanakis *et al.* have reported a method of realizing HMFCF in well studied $Mn_2VZ$ (Z=Al,Si) Heusler compounds by modulating the total number of valance electrons through Co substitution at the Mn site [16]. They show HMFCF in $[Mn_{1-x}Co_x]_2VAl$ when x=0.5 and in $[Mn_{1-x}Co_x]_2VSi$ when x=0.25. This crucial x value corresponds to the total valance electron number of 24. Even though the proposed materials are likely to form since the parent compounds $Mn_2VAl$ and $Co_2VAl$ have been synthesized with $L2_1$ structure, only limited number of reports is available in literature. As far as the films are concerned, Meinert M *et al.* have reported a near compensation of magnetic moment ($M_S$ = 0.1 $\mu_B$/f.u) in $Mn_{2-x}Co_xVAl$ films for x=0.5[17]. Researchers have attempted to synthesize HMFCF in the bulk form by substituting Co at the Mn site of $Mn_2VZ$ (Z=Ga,Al) Heusler compounds. The half-metallicity and ferrimagnetic characteristics of the parent $Mn_2VGa$ compound has been experimentally proven by magnetic, transport and neutron diffraction studies [18, 19]. Ramesh Kumar *et al.* have reported the compensation behavior for $Mn_{2-x}Co_xVGa$ for x=1 for which the valance electrons count is 24 [20]. Investigation carried out by Bhargab Deka *et al.* on the same compound has also indicated the similar observation [21]. But both the investigations have shown a lowering of Curie temperature ($T_C$) with Co substitution at the Mn site. The former group has reported the $T_C$ as 171 K and the later group has reported the $T_C$ as 367 K for x=1 composition. Similar trend in $T_C$ was observed for $Mn_{2-x}Co_xVAl$ also [22]. The transition temperature ($T_C$) decreases from 750 K to 105 K (around 86 % reduction) for 50 % Co substitution in the Mn site of $Mn_2VAl$. As far as the device applications are concerned, the lowering of $T_C$ seems to be a major drawback for the $Mn_2VZ$ (Z=Ga,Al) based HMFCFs. Here we report the interesting observations on the effect of Co substitution at the V site of the half-metallic $Mn_2VZ$ (Z=Ga,Al) compounds which is found to be more efficient in magnetic moment compensation without significantly altering the Curie temperature.

## 2. Experimental details

$Mn_2V_{1-x}Co_xZ$ (Z=Ga, Al and x=0, 0.25, 0.5, 0.75, 1) Heusler compounds have been synthesized by arc melting high pure elements under argon atmosphere. Extra 3% of Mn has been added to compensate the loss. Titanium was melted before the samples to absorb the oxygen left in the chamber. Samples were melted repeatedly after flipping to ensure the homogeneity. After melting, $Mn_2V_{1-x}Co_xGa$ (x=0, 0.25, 0.5, 0.75, 1) samples were sealed inside quartz tubes and annealed for 3 days at 1073 K followed by furnace cooling. $Mn_2V_{1-x}Co_xAl$ (x=0, 0.25, 0.5, 0.75, 1) samples were annealed for 3 days at 673 K followed by furnace cooling . Crystal structure has been determined by x-ray diffraction by using Rigaku smartLab high resolution X-ray diffractometer with Cu-$K_\alpha$ radiation ($\lambda$=1.5418 Å). The Composition was estimated by using energy dispersive X-ray spectrometer attached with FEI-InspectF scanning electron microscope. Low temperature magnetic measurements were carried out using Quantum Design MPMS 3 SQUID VSM and high temperature magnetic measurements were carried out using Lake Shore VSM.

## 3. Results and Discussion

### 3.1 Structural properties

Figure 1a shows the x-ray diffraction (XRD) patterns (using Cu-K$_\alpha$ radiation) of the Mn$_2$V$_{1-x}$Co$_x$Ga compounds with different Co concentration x.

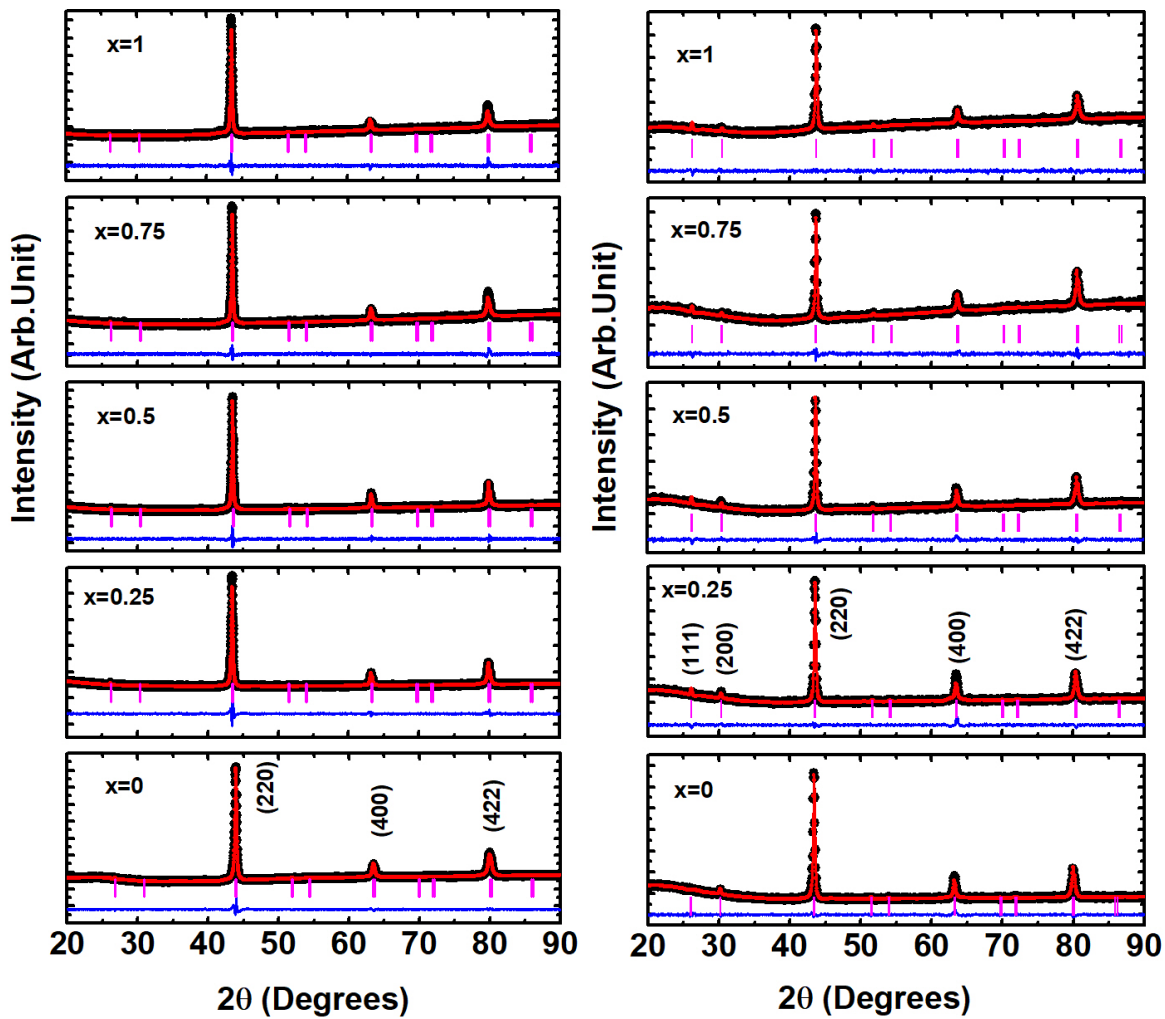

Figure 1a                                    Figure 1b

Figure 1: (a) X-ray diffraction patterns of Mn$_2$V$_{1-x}$Co$_x$Ga (x=0, 0.25, 0.5, 0.75, 1) and (b) Mn$_2$V$_{1-x}$Co$_x$Al (x=0, 0.25, 0.5, 0.75, 1) Heusler compounds measured using Cu-K$_\alpha$ radiation. Plots in black, red, blue and magenta colours represents the experimental data, fitted data after Rietweld refinement, difference between experimental and fitted data and the calculated peak positions respectively.

It is found that the Co substitution has not altered the crystal structure of the parent compound. Only the order independent principal reflections such as (220), (400) and (422) were present in the diffraction pattern and the low angle order dependent superlattice reflections (111) and (200) which are characteristic $L2_1$ (The most ordered crystal structure for the Heusler compounds without atomic anti-site disorder) peaks were absent. One of the possible reasons for the absence of superlattice reflection could be due to the presence of atomic anti-site disorder in the compounds. If the Y and Z atoms in the $X_2YZ$ type Heusler compounds are evenly distributed, the 4a and 4b Wyckoff positions will become equivalent and the resulting disordered structure is known as B2 (space group: 221, B2, Pm-3m). If the X&Y or X&Z atoms are randomly distributed, the resulting structure is referred as $DO_3$ (Space group: 225, $D0_3$, Fm-3m). Among the disordered structures, the most disordered structure is known to be the A2 structure where all the three atoms (X, Y and Z) occupy random positions (space group: 229, A2, Im-3m). The ordered $L2_1$ and disordered B2 and A2 structures with the degree of atomic disorder could be estimated from the XRD pattern. The perfectly ordered $L2_1$ structure would show the low angle order dependent super-lattice reflections (111) and (200) along with the order independent reflections such as the most intense (220), (400) and (422). The structure factor (F) for the super-lattice reflections can be obtained by using equation 1 and 2 given below.

$$F(111) = 4(f_y - f_z) \qquad (1)$$

$$F(200) = 4(2f_x - (f_y + f_z)) \qquad (2)$$

Where $f_x$, $f_y$ and $f_z$ are the average scattering amplitudes from the X,Y and Z sub-lattices respectively. This clearly indicate that if Y and Z atoms are randomly occupied (B2 disorder), then the intensity of the (111) peak would reduce or disappear. The non zero structure factor of (200) peak would still result in the presence of (200) peak even for the B2 disordered structure. But for the A2 disorder, all three scattering terms would become equal causing the absence of both the super-lattice reflections. So in our present investigation, absence of both the superlattice reflections in $Mn_2V_{1-x}Co_xGa$ compounds could be due to the presence of A2 disorder. Another possible reason could be due to the similar atomic scattering factors of the x-ray for the elements Mn,V,Co and Ga which are in the same period in the periodic table. The earlier report on the neutron diffraction studies of $Mn_2VGa$ compound has shown the huge superlattice reflections which was absent in the xrd pattern [19]. This clearly indicates that the compounds in the present investigations are either in ordered $L2_1$ structure or disordered A2 structure. As far as the $Mn_2V_{1-x}Co_xAl$ compounds are concerned, less intense (111) and (200) superlattice reflections are also present along with the order independent principal reflections as shown figure 1b which indicates that the compounds in this series are not in fully disordered structure (A2/B2). To determine the lattice parameter, structural refinement was carried out for all the compounds by using FullProf software assuming the $L2_1$ ordering [23]. Lattice parameter did not follow a linear variation with the Co concentration for the $Mn_2V_{1-x}Co_xGa$ compounds as shown in figure 2. The lattice parameter was found to decrease with the increase in Co concentration up to x=0.25 which is due to the smaller atomic size of Co compared to V. For x>0.25, an increase in lattice parameter was observed up to x=0.75 followed by a decrement for x=1. Similarly for the $Mn_2V_{1-x}Co_xAl$ compounds a non linear decrease in lattice parameter value with Co concentration has been observed as shown in figure 2.

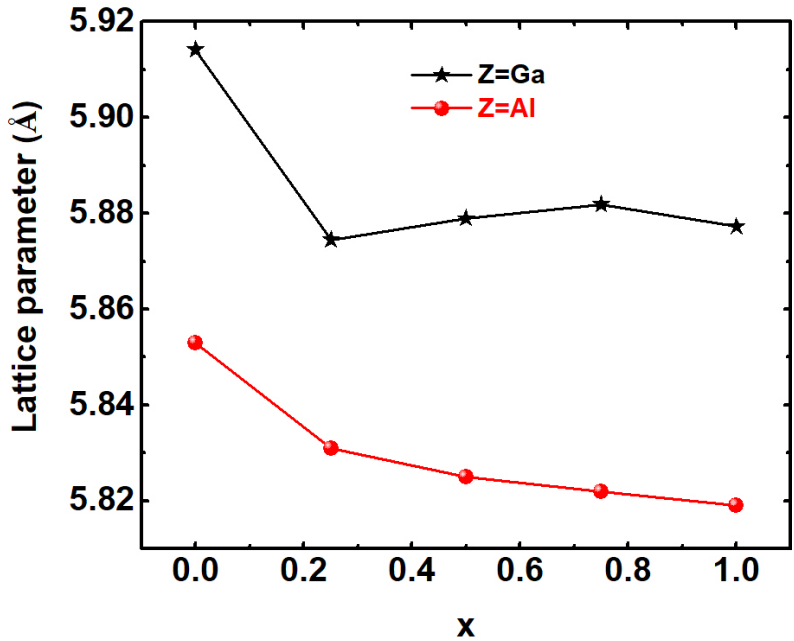

Figure 2: Lattice parameter variation of $Mn_2V_{1-x}Co_xZ$ (Z=Ga,Al;x=0,0.25,0.5,0.75,1) Heusler compounds..Rietveld refinement was carried out by using FullProf software.

The refined lattice parameter values and $\chi^2$ for the fit are compiled in table 1. Back scattered electron (BSE) images have been taken for all the samples to check whether the samples contains any trace of secondary phase. Finely polished bulk samples were used for this purpose. It was confirmed that samples are free from secondary phases.

|   | Mn$_2$V$_{1-x}$Co$_x$Ga | | Mn$_2$V$_{1-x}$Co$_x$Al | |
|---|---|---|---|---|
| x | Lattice parameter | | Lattice parameter | |
|   | a | $\chi^2$ | a | $\chi^2$ |
| 0 | 5.914 | 2.97 | 5.853 | 1.81 |
| 0.25 | 5.874 | 2.39 | 5.831 | 1.66 |
| 0.5 | 5.878 | 1.35 | 5.825 | 1.38 |
| 0.75 | 5.881 | 1.23 | 5.822 | 1.35 |
| 1 | 5.877 | 1.39 | 5.819 | 1.23 |

Table 1: Refined lattice parameter values with respective goodness of fit (Using FullProf software) of Mn$_2$V$_{1-x}$Co$_x$Z (Z=Ga, Al and x=0, 0.25, 0.5, 0.75, 1) Heusler compounds.

## 3.2 Magnetic properties

Since the valance electrons configuration of Mn, V, Ga, Al and Co are $3d^54s^2$, $3d^34s^2$, $4s^24p^1$, $3s^2,3p^1$ and $4s^23d^7$ respectively, Slater-Pauling rule (S-P rule) can be utilized to estimate the total magnetic moment per formula unit ($\mu_B$/f.u) for the Mn$_2$V$_{1-x}$Co$_x$Z (Z=Ga,Al and x=0, 0.25, 0.5, 0.75, 1) compounds. The total number of valance electrons turns out to be 22, 23, 24, 25 and 26 for x=0, 0.25, 0.5, 0.75 and 1 respectively for both series. Thus as per the S-P rule, Mn$_2$V(Ga/Al) should have a magnetic moment of -2 $\mu_B$/f.u and for Mn$_2$Co(Ga/Al), it turns out to be +2 $\mu_B$/f.u. Now for the increase in Co concentration at V site, the moment should decrease to zero for the x=0.5 composition where the total number of valance electrons is the crucial number 24 which constitute the fully compensated magnetic state. It can be calculated that the x=0.25 and x=0.75 composition should have a magnetic moment of -1 and +1 $\mu_B$/f.u respectively. So as per the S-P rule, the magnetic moment of Mn$_2$V$_{1-x}$Co$_x$Z compounds should linearly decrease to zero as x increases from 0 to 0.5 and then raises linearly for higher values of x (x>0.5) reaching the maximum of +2 $\mu_B$/f.u. The negative magnetic moment (calculated from S-P rule) for the composition x<0.5 indicate that the half-metallic band gap would occur in the majority spin sub-band. The measured magnetization curves (M-H curves) at 5 K for the Mn$_2$V$_{1-x}$Co$_x$Ga compounds are shown in figure 3a and figure 3b shows the M-H curves for the $Mn_2V_{1-x}Co_xAl$ compounds which is in good agreement with the above mentioned Slater-Pauling behavior.

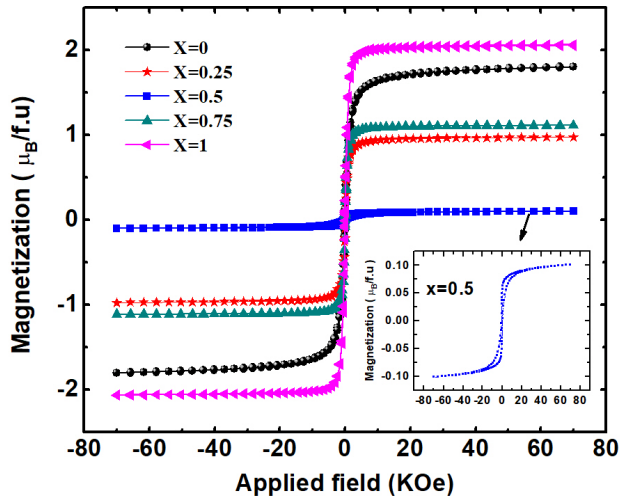 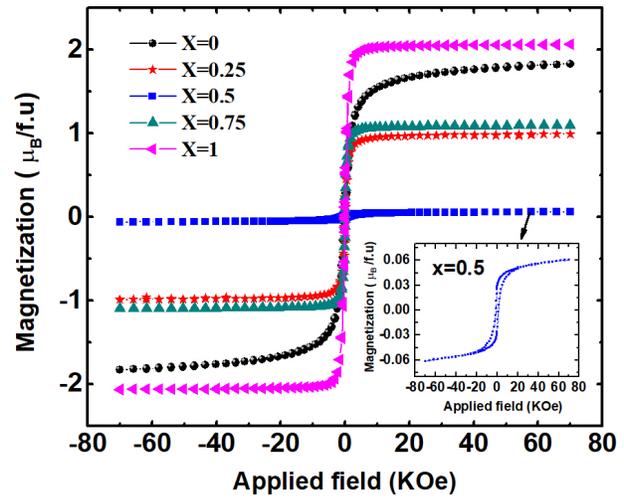

Figure 3a

Figure 3b

Figure 3: (a) Magnetization curves recorded at 5 K for the $Mn_2V_{1-x}Co_xGa$ and (b) $Mn_2V_{1-x}Co_xAl$ compounds. Drastic reduction in magnetic moment was observed for the x=0.5 composition and the corresponding M-H curve alone is shown in the inset.

For a better comparison, the experimental and S-P value of magnetization are plotted together as shown in figure 4 and are complied in table 2.

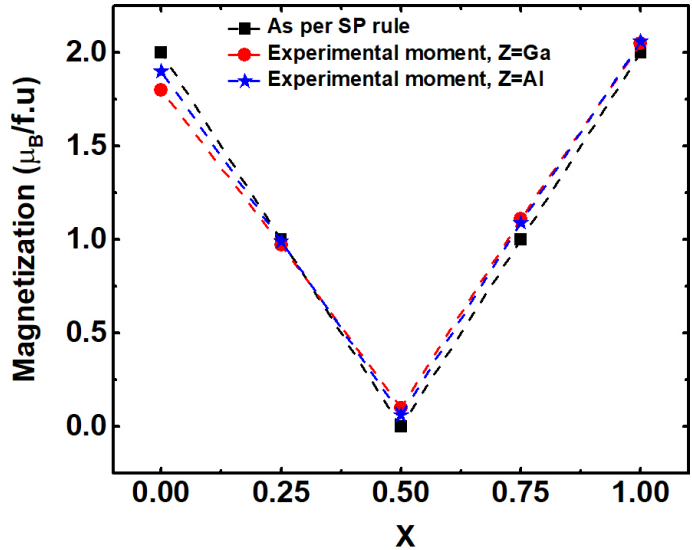

Figure 4: Variation of saturation magnetization with Co concentration for $Mn_2V_{1-x}Co_xZ$ (Z=Ga,Al and x=0, 0.25, 0.5, 0.75, 1) compounds. The values predicted by S-P rule are also plotted together for a better comparison.

| x | Mn$_2$V$_{1-x}$Co$_x$Ga | | | Mn$_2$V$_{1-x}$Co$_x$Al | | |
|---|---|---|---|---|---|---|
| | M$_{S-P}$ ($\mu_B$/f.u) | M$_{Experimental}$ ($\mu_B$/f.u) | T$_C$ (K) | M$_{S-P}$ ($\mu_B$/f.u) | M$_{Experimental}$ ($\mu_B$/f.u) | T$_C$ (K) |
| 0 | -2 | 1.80 | 825 | -2 | 1.83 | 786 |
| 0.25 | -1 | 0.97 | 776 | -1 | 0.99 | 727 |
| 0.5 | 0 | 0.10 | 744 | 0 | 0.06 | 715 |
| 0.75 | +1 | 1.11 | 748 | +1 | 1.09 | 748 |
| 1 | +2 | 2.05 | 774 | +2 | 2.06 | 789 |

Table 2: Theoretical (calculated from S-P rule), experimental magnetic moments and Curie temperatures of Mn$_2$V$_{1-x}$Co$_x$Z (Z=Ga,Al and x=0, 0.25, 0.5, 0.75, 1) Heusler compounds.

Considering the Mn$_2$V$_{1-x}$Co$_x$Ga compounds, the experimental value (S-P value) of magnetization for the x=0, 0.25, 0.5, 0.75 and 1 compositions are 1.80 (-2), 0.97 (-1), 0.10 (0), 1.11(1) and 2.05 (2) $\mu_B$/f.u respectively. It is reported that that the small amount of atomic anti-site disorder between V and Ga atoms results in the lowering of magnetic moment of parent Mn$_2$VGa. Our value of 1.80 $\mu_B$/f.u is comparable with the earlier reported values of 1.85 $\mu_B$/f.u [19]. The small deviation of experimental magnetic moments from the S-P value could be due to the atomic anti-site disorder present in the system. Since a perfect composition is required to obtain the magnetization value predicted by the S-P rule, the small deviation of our experimental values may be due to small compositional variation also. Earlier reports on the effect of Co substitution at the Mn site also have shown the compensation behavior at x=1 for the Mn$_{2-x}$Co$_x$VGa compound. Ramesh Kumar et al. have reported a magnetic moment value of 0.37 $\mu_B$/f.u for (MnCo)VGa which should have exhibited a zero magnetic moment according to the S-P rule. Their magnetic moment values of the compounds did not follow the S-P rule indicating deviation from half-metallicity. Here it is to be highlighted that their investigation has revealed a lowering of T$_C$ value with Co substitution at the Mn site. T$_C$ has reduced to 171 K for the x=1 composition which is very low compared to that of the T$_C$ of parent Mn$_2$VGa compound (783 K). This would be a major drawback as far as the practical applications are considered. Similar trend in magnetic moment compensation and T$_C$ as a result of Co substitution at the Mn site was observed by Bhargab Deka et al. for the same compound.

They reported magnetic moment value of 0.3 $\mu_B$/f.u and $T_C$ of 367 K for the (MnCo)VGa compound. The neutron diffraction measurements carried out by Ramesh Kumar *et al.* have revealed the existence of ferrimagnetic order in $Mn_2VGa$ compound having the magnetic moment value of 1.28 and -0.71 $\mu_B$ for Mn and V atoms respectively. The Mn atoms in the different sub lattices are ferromagnetically coupled with each other with equal magnetic moment values. Considering the total magnetic moment contribution arising from two Mn atoms and one V atom, the total magnetic moment per formula unit has turned out to be 1.85 $\mu_B$/f.u which is in very good agreement with our experimental value of 1.8 $\mu_B$/f.u in the present investigation. When the Co atom is substituted at the Mn site, it ferromagnetically couples with the V moment and this effective moment is antiparallel to the Mn moment, resulting in a reduction in the total magnetic moment for the $Mn_{2-x}Co_xVGa$ compounds. The reported compensated moment value was in the range 0.3 to 0.37 $\mu_B$/f.u for the x=1 composition by Ramesh Kumar *et al.* [20] and Bhargab Deka *et al.* [21]. We believe that a similar compensation is happening for the V site substitution also. Even though the magnetic moments of Mn, V and Co atoms in $Mn_2V_{1-x}Co_xGa$ compounds are aligned as in the case of Mn site substitution, the exact reason for the compensation could be slightly different since we are not removing any Mn atoms here. Since the Mn and V moments are anti-parallel, either the moment of Mn has to decrease or the effective moment originating from V and Co atoms has to increase for the reduction in total magnetic moment. First principle calculation showed that for the Mn site substitution, as the Co concentration was increased, the magnetic moment of Mn and V atoms tend to decrease and for the Co atom it increases since the exchange interaction between Co-Co was getting stronger [24]. Since the substituted Co atoms can sit in any of the eight Mn sites randomly (considering the super cell), the Co-Co inter-atomic distance for the Mn site substitution could be more compared to the Co-Co distance for the V site substitution. So possibly the effective Co moment in the V site substituted compounds could be higher than that of Mn site substitution resulting in a better compensation of magnetic moment. Now considering the $Mn_2V_{1-x}Co_xAl$ compounds, the experimental value (S-P value) of magnetization for the x=0, 0.25, 0.5, 0.75 and 1 compositions are 1.83(-2), 0.99 (-1), 0.06 (0), 1.09(1) and 2.06 (2) $\mu_B$/f.u respectively. Here also a close matching between the experimental and S-P rule predicted magnetization values are observed as shown in figure 4. The explanation for the magnetic moment compensation holds the same as that of the $Mn_2V_{1-x}Co_xGa$ compounds.

As mentioned before, the major drawback of the earlier reported Mn site substitution was the lowering of $T_C$ as the compensation point approaches. Interestingly our investigation on the Co substitution at the V site of $Mn_2VZ$ (Z=Ga, Al) compounds reveals nearly full compensation of magnetic moment without a significant reduction in $T_C$. We have achieved a compensated magnetic moment value of 0.1 $\mu_B$/f.u and $T_C$ of 744 K for the $Mn_2V_{0.5}Co_{0.5}Ga$ compound and for $Mn_2V_{0.5}Co_{0.5}Al$ it turned out to be 0.06 $\mu_B$/f.u and 715 K. Theoretical investigation carried out by E. Şaşıoğlu *et al.* have shown that the Mn-V ferrimagnetic interaction is almost five times stronger than the Mn-Mn interaction for the Mn$_2$VAl compound and hence the Curie temperature is mostly influenced by Mn-V interaction [25]. So when the Mn atoms are substituted by the Co atoms, the exchange interaction between Mn and V becomes weaker and as a result, the T$_C$ tends to decrease. On the application point of view, this could be a disadvantage since the thermal stability of the spintronic devices are one of the important criterions. On the other hand, our observations on the V site substitution are in contrast to the results of Mn site substitution. Figure 5a and figure 5b shows the temperature variation of magnetization for the Mn$_2$V$_{1-x}$Co$_x$Z (Z= Ga, Al and x=0, 0.25, 0.5, 0.75, 1) Heusler compounds.

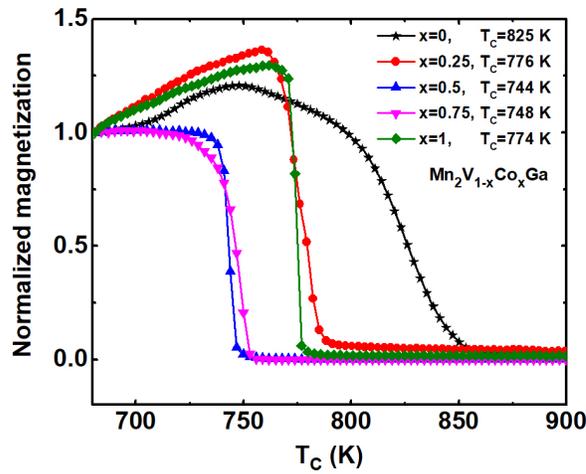 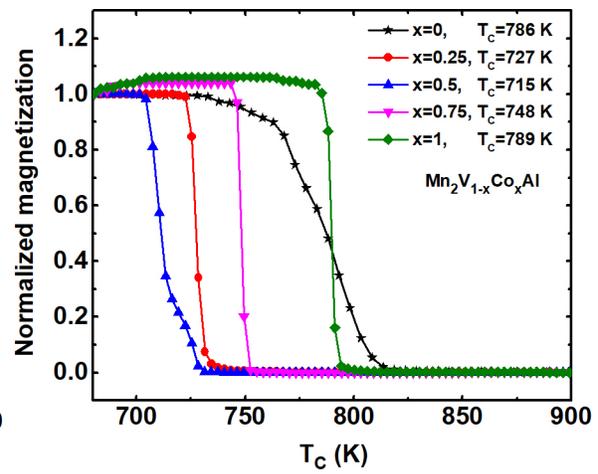

**Figure 5a**  **Figure 5b**

Figure 5: (a) M-T curves for the Mn$_2$V$_{1-x}$Co$_x$Ga and (b) Mn$_2$V$_{1-x}$Co$_x$Al compounds. T$_C$ was estimated by taking the first derivative of the M-T curves. T$_C$ was more than 700 K for all the compounds which is in sharp contrast to the earlier reports on Mn site Co substitution.

Considering the Mn$_2$V$_{1-x}$Co$_x$Ga compounds, the obtained T$_C$ for the Mn$_2$VGa compound is 825 K which is close to the earlier reported values [19, 20]. As the Co content was increased, T$_C$ was decreased to 776 K for x=0.25 and for x= 0.5 a further drop to 744 K was observed. As x reaches 0.75, T$_C$ started increasing and reaches a slightly higher value of 748 K and for Mn$_2$CoGa it increases to 774 K. The variation of T$_C$ with Co concentration is shown in figure 6. Note that compared to the parent compound's T$_C$, only a small reduction in T$_C$ was observed (9.8 % for x=0.5) for the V site substituted compounds which is in sharp contrast to the earlier reports on the Mn site substitution (78 % [20] and 52 % [21] reduction for x=1 composition). Now considering the $Mn_2V_{1-x}Co_xAl$ compounds the obtained $T_C$ for the parent $Mn_2VAl$ and $Mn_2CoAl$ are 786 K and 789 K respectively. $T_C$ was found to decrease with x=0.25 and 0.5 followed a by an increase for x=0.75 and 1 as shown in figure 6. Here also compared to the $T_C$ of the parent compound, only a small reduction in $T_C$ was observed (9 % for x=0.5) for the V site substituted compounds which is in sharp contrast to the earlier reports on the Mn site substitution (86 % reduction for MnCoVAl compound).

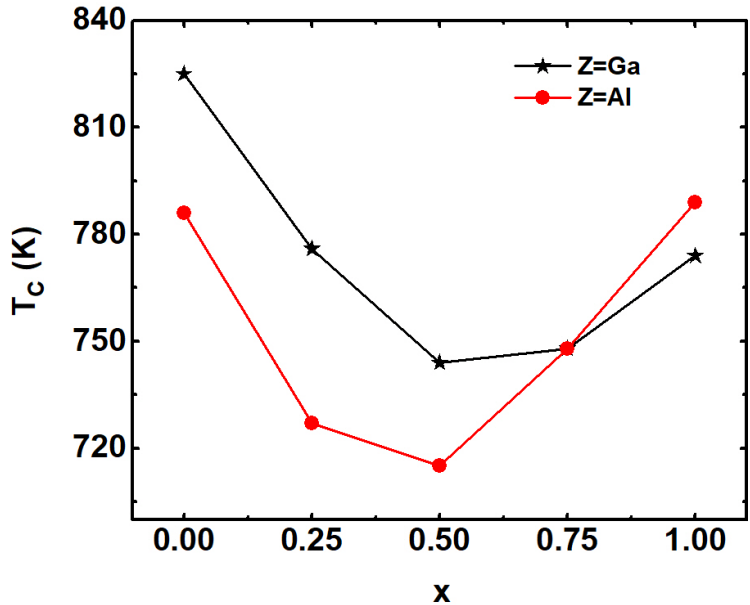

Figure 6: Variation of $T_C$ for the $Mn_2V_{1-x}Co_xZ$ (Z=Ga,Al and x=0, 0.25, 0.5, 0.75, 1) Heusler compounds.

The high $T_C$ behavior can be understood as follows. For $Mn_2VAl$ the major contribution to $T_C$ arises from the Mn-V interaction since it is much stronger than the Mn-Mn and V-V interaction as mentioned earlier. The same could be the case with the isostructural $Mn_2VGa$ compound. When Co is substituted at the Mn site, it would have weakened the Mn-V exchange interaction and thereby causing the reduction in $T_C$. The observed nearly equal values of $T_C$ for $Mn_2VGa$ (825 K) and $Mn_2CoGa$ (774 K) in the present study indicate that the Mn-V and Mn-Co exchange interaction strength in the respective compounds may not be very different and when we substitute Co at the V site, the Mn-Y (V/Co) exchange interaction which contribute to the major fraction of the $T_C$ would not have weakened. The same explanation could be applicable to Co substituted $Mn_2VAl$ also.

## 4. Conclusions

Nearly full compensation of magnetic moment was achieved for 50 % Co substitution at the V site of $Mn_2VZ$ (Z=Ga,Al) Heusler compounds without altering the Curie temperature significantly. The compensated magnetic moment for $Mn_2V_{0.5}Co_{0.5}Z$ compounds are 0.1 $\mu_{B/f.u}$ for Z=Ga and 0.06 $\mu_B$/ f.u for Z=Al which is a 95 % and 97 % reduction from that of the parent compounds $Mn_2VGa$ and $Mn_2VAl$ respectively. The transition temperature ($T_C$) decreases from 825 K to 744 K (only 9.8 % reduction) for 50 % Co substitution in the V site of $Mn_2VGa$ which is in sharp contrast to the earlier reported drastic reduction in $T_C$ (around 52 % to 78 %) observed for Mn site substitution of Co. For $Mn_2V_{0.5}Co_{0.5}Al$, only 9 % reduction of $T_C$ was observed compared to $Mn_2VAl$ which is far better than the reported 86 % reduction for the Mn site substituted compound. The experimentally determined magnetic moment values of all the compounds follows the Slater-Pauling behavior expected for half-metallic full-Heusler compounds. Our investigation shows that the Co substitution at the V site is better than the substitution at the Mn site of $Mn_2VZ$ (Z=Ga, Al) Heusler compounds in terms of compensated magnetic moment value and Curie temperature. The achieved stable $T_C$ well above room temperature in the present study would avoid the practical difficulty in fabricating spintronic devices utilizing the $Mn_2VZ$ (Z=Ga, Al) based fully compensated half-metallic ferrimagnets.